\documentclass[letterpaper, 10pt, conference]{ieeeconf}  % Comment this line out
\usepackage{graphicx}
\usepackage{amsmath}
\usepackage{amssymb}
\usepackage{graphics} % for pdf, bitmapped graphics files

\title{\LARGE \bf
Effect of particle size and interparticle spacing
on dislocation behavior on Nickel based super-alloys
}

\author{Dr.Anand Krishna Kanjarla*,Gautham Muthusamy(MM13B040)*, Aditya Venkatraman(ME13B102)**}

\begin{document}

\maketitle
\thispagestyle{empty}
\pagestyle{empty}

%%%%%%%%%%%%%%%%%%%%%%%%%%%%%%%%%%%%%%%%%%%%%%%%%%%%%%%%%%%%%%%%%%%%%%%%%%%%%%%%
\begin{abstract}

Ni-based superalloys have been the subject of enormous usage in scenarios where the loading is heavy and often occurs at elevated temperatures. The strengthening mechanisms that come into play within the metallic lattice have been studied extensively as micromechanical MMC models. These continuum formulations suffer from several limitations. The underlying mechanisms at the atomistic scale have not yet been well understood. The report attempts to model the interaction of moving dislocation with cuboidal precipitates and explain the strengthening effect. The effect of particle size and inter-particle distance on the strength are evaluated. Several physically meaningful results have also been interpreted and shown.

\end{abstract}

%%%%%%%%%%%%%%%%%%%%%%%%%%%%%%%%%%%%%%%%%%%%%%%%%%%%%%%%%%%%%%%%%%%%%%%%%%%%%%%%
\section{INTRODUCTION}
\subsection{Motivation}

Nickel-based Superalloys are found in a wide range of applications. The most
prominent use is in the manufacture of gas turbines for use in commercial and
military aircraft, power generation, and marine propulsion.A jet engine experiences temperatures ranging from 300K to 1500K along with very high pressure. Superalloys exhibit high creep resistance
at high temperatures, good surface stability, and corrosion and oxidation
resistance.Plastic deformation of any material can be understood by scaling
down to the molecular level.At this scale,the movement of dislocations play a vital role in controlling the plastic deformation. The distribution and size of precipitates Ni$_{3}$Al acts as a obstruction to the movement of dislocations.

\subsection{Theory}

The microstructure of Nickel-based Superalloys consists of
ordered $\gamma'$ Ni$_{3}$Al precipitates with a L1$_{2}$ structure, coherently
set in the $\gamma$-matrix, a face-centred cubic (fcc) nickel-based solid
solution. Nickel-based superalloys derive much of their excellent mechanical properties at high temperature from the $\gamma'$ precipitates. They have a roughly cuboidal shape.The dimensions of precipitates and channels are in the sub-micrometer range.The essential role of $\gamma'$ precipitate in the matrix is to obstruct the movement of dislocations by acting as a bulk obstacle.The $\gamma$-$\gamma'$ interface plays a vital role as there exists a misfit between
the lattice parameter of the precipitate and the matrix. Lattice misfit is a measure of the incoherency between the $\gamma'$ precipitate and the
$\gamma$ matrix.Mathematically,misfit can be calculated as
\[\delta = 2\frac{a_{\gamma'} - a_\gamma}{a_{\gamma'} + a_\gamma}\]
where, a$_\gamma$ is the lattice parameter of the $\gamma$ matrix and a$_{\gamma'}$ is the lattice parameter of the $\gamma'$ precipitate.The lattice misfit that is specified in a given simulation affects the overall dynamics of dislocations observed and subsequently the prediction of the properties. From the scientific literature, the values of a$_\gamma$ and a$_{\gamma'}$ are found as 3.52\AA and 3.529\AA  respectively.The misfit is hence very small and assumed to be coherent in the study.

\subsection{Objective}
The objective of the project is to perform a Molecular Dynamics simulation to study the interaction between the precipitate and the interaction.One of the important goals of the project is to investigate the dislocation precipitate interaction for different precipitate sizes and distributions.In order to achieve this, a model has to be setup to perform the simulation.The system consists of a Nickel ($\gamma$) matrix ,a single edge dislocation and two cuboidal Ni$_{3}$Al ($\gamma'$) precipitates.

This paper is organized as follows. Section 2 provides a brief literature survey.Section 3 describes the procedure to setup the simulation cell,parameters associated the model.Section 4 shows the results of the simulation, observations and conclusion.Section 5 is used for acknowledgements

\section{Literature Review}

\subsection{Ni-based Superalloys}

%%%%%%%%%%%%%%%%%%%%%%%%%%%%%%%%%%%%%%%%%%%%%%%%%%%%%%%%%%%%%%%%%%%%%%%%%%%%%%%%

Ni-based superalloys are very useful for high-temperature applications because of their strength retention capabilities and relatively high creep compliance. The shape of the precipitates (Ni$_{3}$Al) is usually in the form of thin plate-like structure or cuboidal when the fraction of the volume occupied by the precipitate is moderately high. The interaction of dislocations with precipitates can result in climb, bow-out (the so-called orowan bowing) or it could result in the cutting of the precipitate by the dislocation
While precipitation hardening has been studied extensively using continuum mechanics, its underlying mechanism in the atomistic scale has not been investigated satisfactorily. The continuum assumption makes use of both long-range and short range forces (ie), those that arise from the intrinsic stress field associated with a screw/edge dislocation or those that arise from the interaction of the dislocation and the particle, which are in turn, thought to originate from the mismatch of the elastic properties between the matrix and precipitate phases. Eshelby's equivalence principle has been used in various forms to achieve homogenization of the precipitate hardened alloys to simulate their macroscale response. There are several major assumptions made by these analytical constitutive methods that attempt to study dislocation particle interaction. The assumptions include, but are not limited to line tension in the dislocation line, coherency between precipitate and the matrix phase, and relatively smooth dislocation lines. They completely negate the possiblity of formation of stacking fault which would lead to the introduction of anisotropy in an otherwise isotropic polycrystal. The smoothness of dislocation is another gross-simplification that negates the possibility of the formation of kinks around the precipitate, especially if the precipitate is cuboidal. The strengthening effect is particularly pronounced if the volume fraction of the precipitates is moderately high, as is the case with stable cuboidal precipitates, generally.

\subsection{A brief look at continuum based approaches}

The first ever publication on the continuum mechanics based interaction between inclusions and dislocations was published by Nabarro (1940)\cite{Mott1940}. He calculated the stresses in a matrix with a harder inclusion based on difference in lattice parameters. The internal force attributed to the system is non-zero even in the absence of external loading due to the intrinsic stress-field around a dislocation. Ardell\cite{ardell1985precipitation} and Ashby provide greater detail in their discussions about the interactions of precipitates and matrix. These discssions, are again based on the line-tension concept. The discussion is also limited by its application to only dispersed particulates in an infinite medium. The effect of Matrix non-linearity and aspect ratio  on strength are studied in Lee and Mear (1991)\cite{lee1991effect}. A energy based formulation was developed by Zhu and Zbib (1995)\cite{zhu1995macroscopic} to study the elasto-plastic response of MMC to understand the strengthening effect. They modeled the MMC as rigid inclusions in an infinite plsatically flowing matrix. An energy-based framework was constructed from which the constitutive model was derived.
 As discuused before, the continuum models are restricted to specific geometries and are not generic formulations that can be used to address complicated loading scenarios.
 \section{Usage of molecular dynamics }
 The short-range interactions can be effectively studied with the LAMMPS molecular dynamics package. This provides us with a clear understanding of the unit processes that are
activated when a dislocation interacts with a given precipitate. The system studied here is  the coherent Ni-Ni$_{3}$Al matrix precipitate super-alloy. 
\subsection{Model definition and lattice generation}
The geometry of the system consists of the Nickel matrix and the Ni$_{3}$Al precipitate. The system consists of cuboidal precipitates in a single crystal of Nickel. The orientation is along the cartesian axes, because the stacking fault energy of Nickel is particularly low on the [100] direction. This is proven to be unneccessary as the formation of a dislocation loop which always happens when the dislocation starts bowing around the precipitate.  The alloy is created using a  user-defined basis for a crystal structure that contains Al at the corners and Ni atoms at the face centres, as is the case with the precipitate phase. The basis allocates a particular type of atom to each position and generates the lattice along the specified orthonormal directions. 
\subsection{Creation of the dislocation}
Two methods were devised to create the dislocations, as explained in [6]. The first one involoved creation of known sources of dislocations, like voids. The major disadvantage with this setup was the fact that voids only created partials, whereas for our simulations, we needed full edge dislocations. This was overcome by using the other method suggested in [6], which was, define the basis associated with the lattice (both precipitate and matrix), and dump the coordinates to a file. The coordinates were read from this file in a third party application called Atomsk which allowed us to delete a set of half-planes and move the remaining atoms according to the linear elastic displacement field surrounding the dislocation (in this case, an edge dislocation). One other way of creating a dislocation in LAMMPS itself was to remove two half planes and let the system go to equilibrium, thereby evolving itself into a system of two partials that can be controlled using shear forces. However, this method was not adopted after significant proliferation due to the enormous times involved in equilibriation within which partials and misfits associated with the precipitate-matrix phase started evolving, hence undercutting the purpose of the study. As it was mentioned before, the coordinates of the lattice without the dislocation is dumped into a file which is read by atomsk. The dislocation line direction and glide plane are taken as inputs, along with the magnitude of the burgers vector and the poisson's ratio for the matrix. The output created by these arguments is used in the analysis.
\begin{center}
Fig1. An edge dislocation created in Atomsk
\includegraphics[width=5cm]{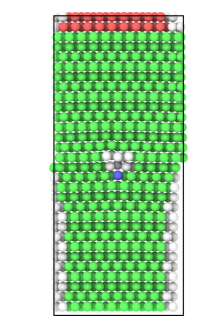}

\end{center}
\begin{center}

\includegraphics[width=5cm]{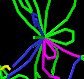}

Fig (2) - A misfit at the centre of the precipitate

\end{center}

\subsection{Parameters associated with the simulation}

The compute style command are used to calculate the thermodynamic outputs of the system. By default, temperature pressure and volume at the end of the current timestep are provided as outputs. However, other parameters required to study the effects of dislocation motion, such as stress, volumetric deformation, strain are required to be input by the user.
 The parameters calculated here include stress/atom and voronoi/atom, which is in-turn used to come up with a volume averaged stress measure, over all the voronoi cells. Following the specification of the compute command, the energy minimization is done with conjugate gradient method. The potential used is Mishin's EAM potential for Ni-Al alloy. The EAM potentials are calculated as follows.
 
 \hspace{30pt} \boldmath{$U_{i}=F_{\alpha}\left(\underset{i\neq j}{\sum}\rho_{\alpha\beta}\left(r_{ij}\right)\right)+\frac{1}{2}\underset{i\neq j}{\sum}\phi_{\alpha\beta}\left(r_{ij}\right)$}
 
 An energy tolerance of 1 pJ was adopted. Note that this isn't particularly low (usually, the values provided are in the order of fJ). This is because, the linear elastic displacement field associated with the dislocation is already programmed into the system. Also, reducing the energy tolerance caused the computational time required to increase dramatically, so it was increased. The misfits associated with the Ni-Ni$_{3}$Al were disappearing and since they are critical to studying the mechanisms in the atomistic scale, the energy tolerance was increased to 1 pJ. It is unclear why this phenomenon is observed. Further investigation is necessary here. 
 Following the energy minimization, an ensemble of velocities are created to achieve equilibriation via temperature rescaling at a set number of NVE integration timesteps. Note that the window within rescaling is done and fraction to which it is done in this simulation is set at 10.0 and 1.0 (1.0 meaning that temperature is reset to the system temperature/desired equilibrium temperature), respectively, which acknowledges the fact that temperature deviations (even slight ones from set temperature (in this case, 30K), strongly affect properties like intrinsic dislocation densities and especially stacking fault energies. The thermal temperatures are calculated after subtracting explicitly the bulk advection components using the keyword "thermo-modify". The equilibriation was performed for 5ps and the temperature results shown in the corresponding section of the paper. 
 The shearing action is accomplished by setting the velocity boundary condition on the top surface. This is followed by velocity rescaling at every timestep to ensure equilibriation at every 10 timesteps. The temperature is re-calculated taking into consideration only the glide direction and dislocation direction as velocity components. Thus this option overrides the default method for computing Temperature, for the purposes of this analysis.
 The values of stress/atom and voronoi/atom obtained from this analysis is dumped into a lammps dump file, which is later used for visualiation with Ovito.

 \begin{center}

\includegraphics[width=8cm]{Untitled_Diagram.png}
\newline
Fig (3)- Simulation flow chart
\newline
\newline
\end{center}
 \subsection{Model parameters}
 The dimensions and the model parameters are shown below. The distance and the particle size are subject to variation and are denoted by 'd' and 's' respectively. 
The parameter d was varied from 20 lattice units to 50 lattice units, in steps of 10, whereas 's' was taken as 10,15 or 20 lattice units.
\begin{center}
\includegraphics[width=9cm]{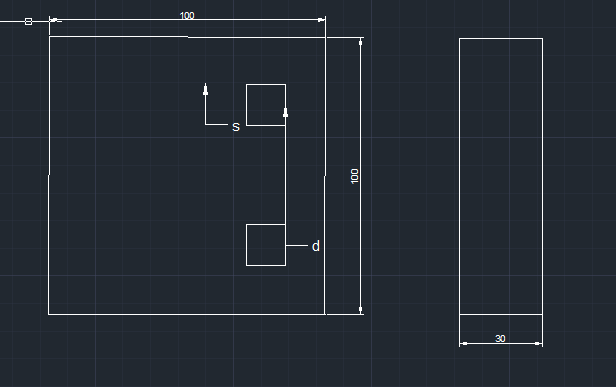}
Fig (4) - Diagram showing the model dimensions
\newline
\end{center}
\section{SIMULATION RESULTS}
\subsection{Qualitative and physically meaningful results}
  The energy minimization results in a lattice with the two partials that were explicitly created with Atomsk as well as misfits associated with matrix-precipitate mismatch.
\begin{center}
\includegraphics[width=8cm]{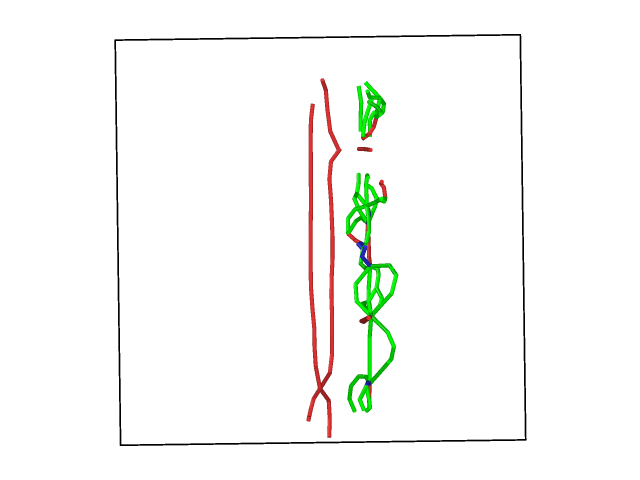}
\linebreak
Fig (5)-Showing initial dislocations 
\end{center}

The figure contains several shockley dislocations with burgers vector 1/6$[112]$. The general motion of the dislocations and their interactions with the precipitates consists of distinctly observable regimes. It first starts with the dislocation hitting the precipitate phase. The force acting on the dislocation due to the applied shear stress is always perpendicular to the burgers vector and dislocation line between the two precipitate phases will bow out and can generate dislocation loops. The normal force couple created due to the shearing, acting on the dislocation will produce a torque on the dislocation line between the precipitate phases along a direction perpendicular to the slip plane. As a result, dislocation line first assumes a semi-circle and with force per unit length acting normal to the line vector the dislocation line then assumes a "bow" shape and a dislocation loop is generated. One must note that for a particular burgers vector of the dislocation, the burgers vector of the misfit generated due to the misfit between the matrix and the precipitate will have a particular orientation, as discussed above. The stacking fault is present within the dislocation loop and its crystal structure is HCP, as expected. 
Note that this rendered picture (fig (7)) gives the impression of two different dislocations interacting. However, this is a periodic cell, implying that the dislocation in this cell is interacting with the bow-out from the previous cell, thereby giving the appearance of two partials from the same cell interacting with each other. The misfit dislocations at the particle act as effective pinning agents for dislocation motion, thereby hardening the alloy.

\begin{center}
\includegraphics[width=8cm]{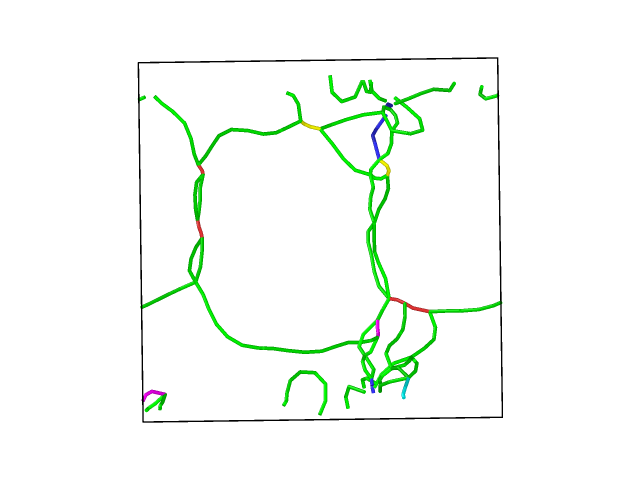}
\linebreak
Fig (6) - Dislocation bow-out around the particle
\end{center}
The stacking fault generation is expected wherever there are dislocation loops. One of the loops is found in between the two particles, in the y-z direction (not shown). This is shown in the figure below.
\begin{center}
\includegraphics[width=8cm]{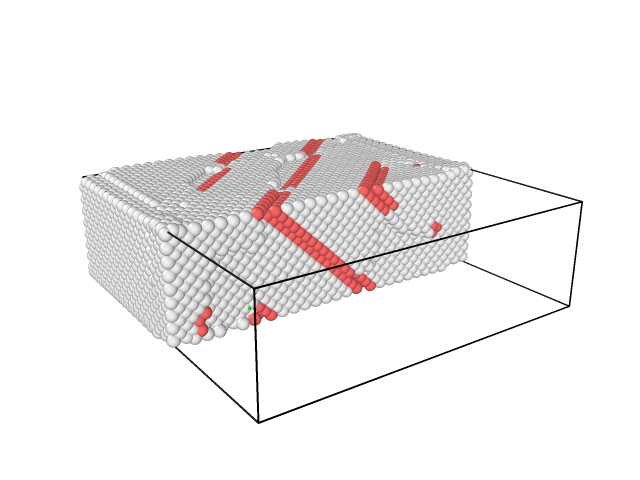}
\linebreak
Fig (7) - Stacking fault showing HCP Ni around the particle
\newline
\newline
\newline
\newline
\newline
\newline
\newline
\newline
\end{center}
This is swiftly followed by the annihilation of the partials with interacting dislocations, be it the misfit shockleys or the next periodic cells edge dislocation. This leaves only the misfits in the lattice, as shown. Some strays are bound to be leftover because the periodicity is lost in [010] direction. A definite drop in the stress is expected due to this phenomenon.
Further shearing causes further generation and destruction of dislocation causing other bumps in the stress/strain plot.
\newline
\newline
\newline
\newline
\newline
\newline
\newline

\begin{center}
\includegraphics[width=8cm]{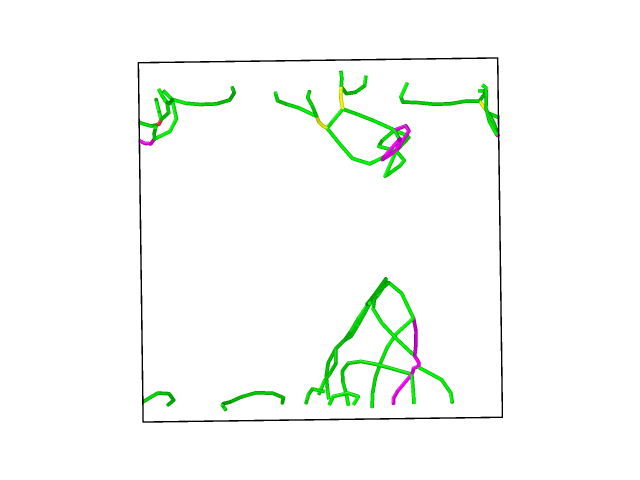}
\linebreak
 Fig (8) - Lattice with only misfits
\end{center}
\subsection{Qualitative results}

\begin{center}
\includegraphics[width=8cm]{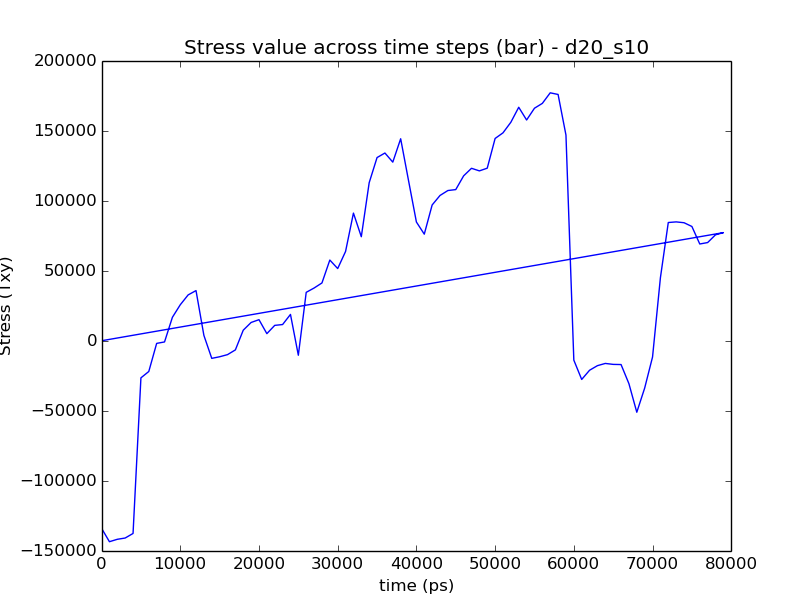}
\linebreak
Figure(9) - Stress vs Time step plot for precipitate distance of 20 lattice units and size of precipitate 10 lattice units.
\end{center}

\begin{center}
\includegraphics[width=8cm]{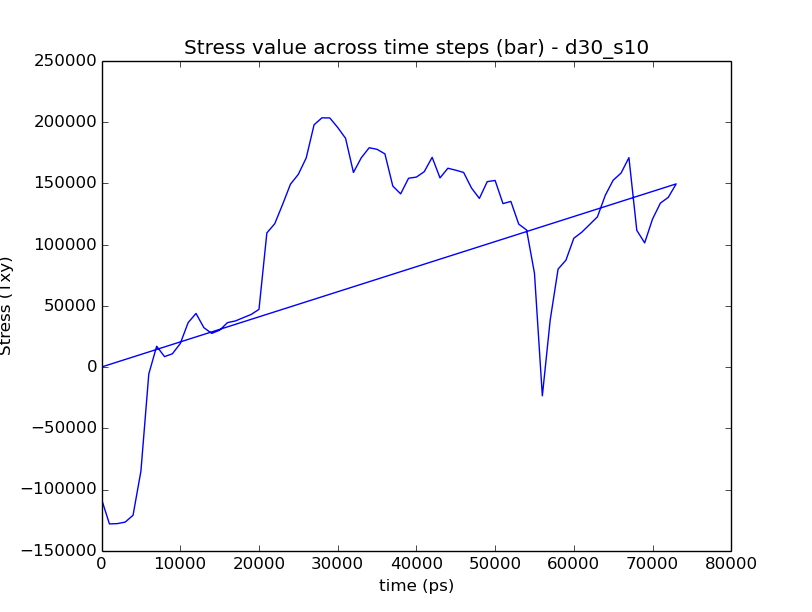}
\linebreak
Figure(10) - Stress vs Time step plot for precipitate distance of 30 lattice units and size of precipitate 10 lattice units.
\end{center}

\begin{center}
\includegraphics[width=8cm]{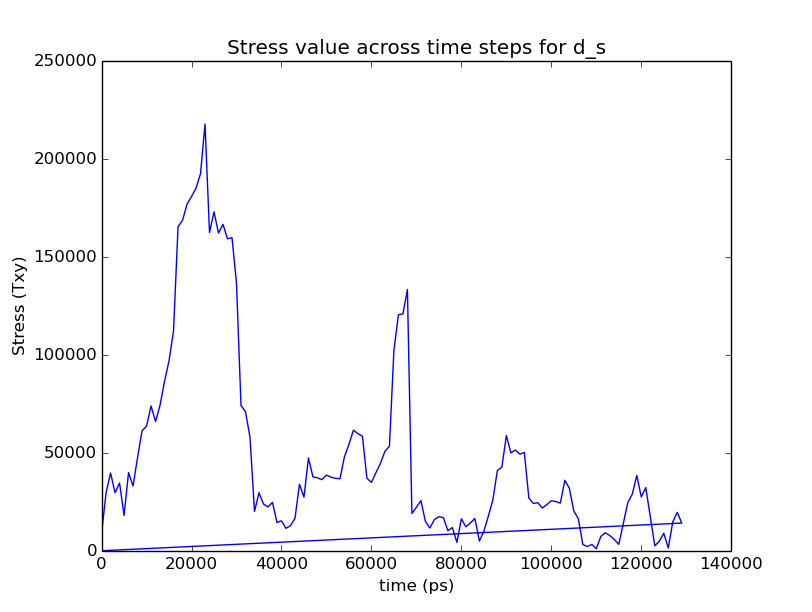}
\linebreak
Figure(11) - Stress vs Time step plot for precipitate distance of 30 lattice units and size of precipitate 20 lattice units.
\end{center}

\begin{center}
\includegraphics[width=8cm]{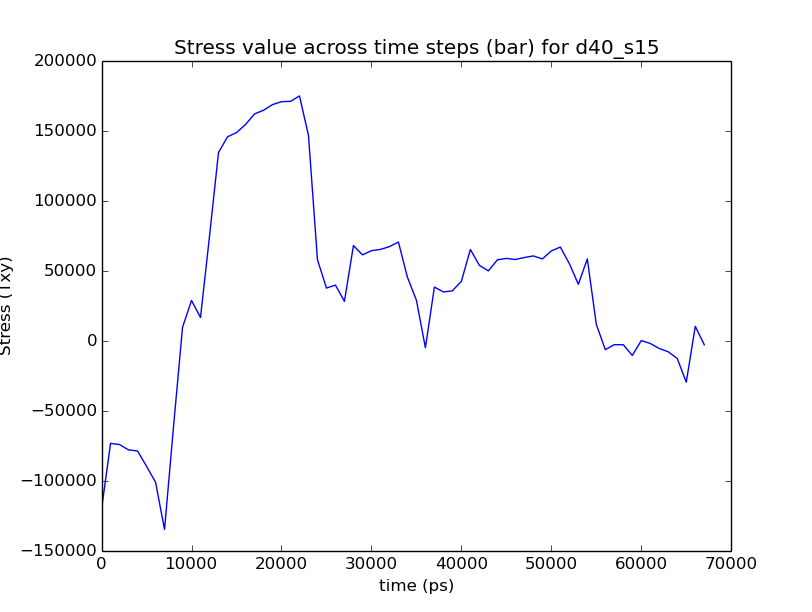}
\linebreak
Figure(12) - Stress vs Time step plot for precipitate distance of 40 lattice units and size of precipitate 15 lattice units.
\end{center}

\begin{center}
\includegraphics[width=8cm]{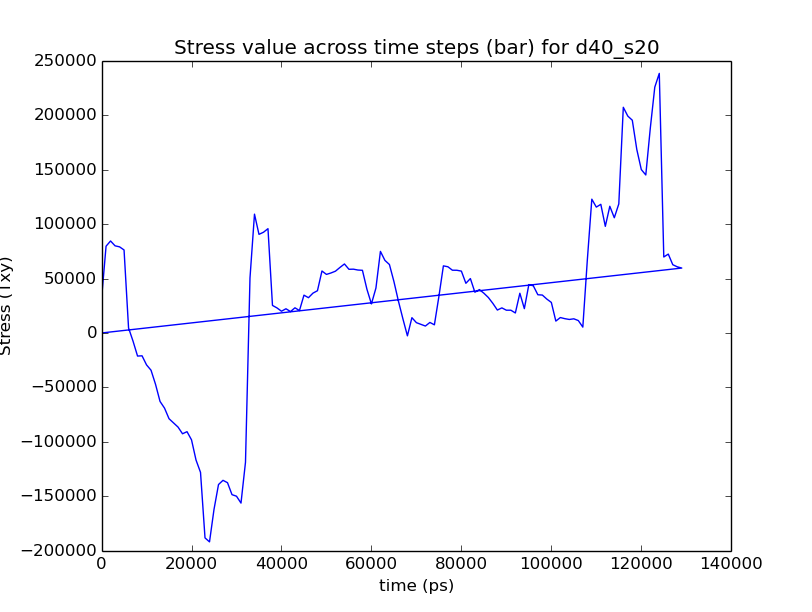}
\linebreak
Figure(13) - Stress vs Time step plot for precipitate distance of 40 lattice units and size of precipitate 20 lattice units.
\end{center}

\begin{center}
\includegraphics[width=8cm]{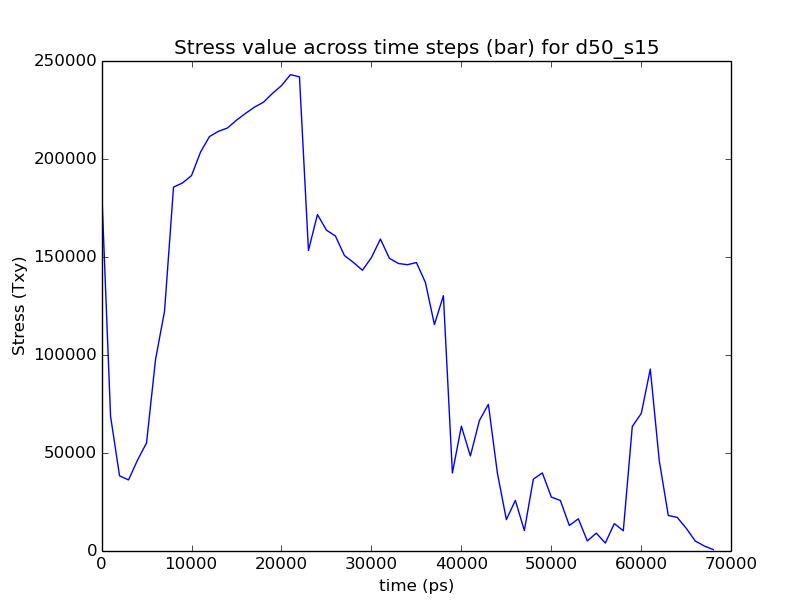}
\linebreak
Figure(14) - Stress vs Time step plot for precipitate distance of 50 lattice units and size of precipitate 15 lattice units.
\end{center}
As explained before, the strength is a strong function of the size and inter-particle distance. Here, it is tabulated for a few sample cases to illustrate the contrast.
\begin{table}[h]
\caption{VARIATION OF MAXIMUM STRESS}
\label{table_example}
\begin{center}
\begin{tabular}{|c||c||c||c|}
\hline
Distance(lattice units) & Size & Max Stress (bar) & Timestep\\
\hline
10 & 10  & 106137.2402539 & 26000\\
\hline
20 	& 10 &  177092.09471901 & 57000\\
\hline
30 	& 10 &  217729.24225723 & 23000\\
\hline
40 	& 15 &  174884.42767411 & 22000\\
\hline
40 	& 20 &  238394.39515007 & 124000\\
\hline
50 & 15 &  242913.97642214 & 21000\\
\hline
50 	& 20 &  249883.47839079 & 56000\\
\hline
\end{tabular}
\end{center}
\end{table}

As is evident from the table, the strength is directly proportional to both the inter-particle distance as well as particle size $[with the outlier simulation being d40, s15]$

The thermos output of temperature was read from the log file and it was plotted to ensure system was at equilibrium at the start of the verlet run. The plot for one of the simulation cases is shown. There hasn't been much of a difference between the various scenarios. 
\begin{center}
\includegraphics[width=8cm]{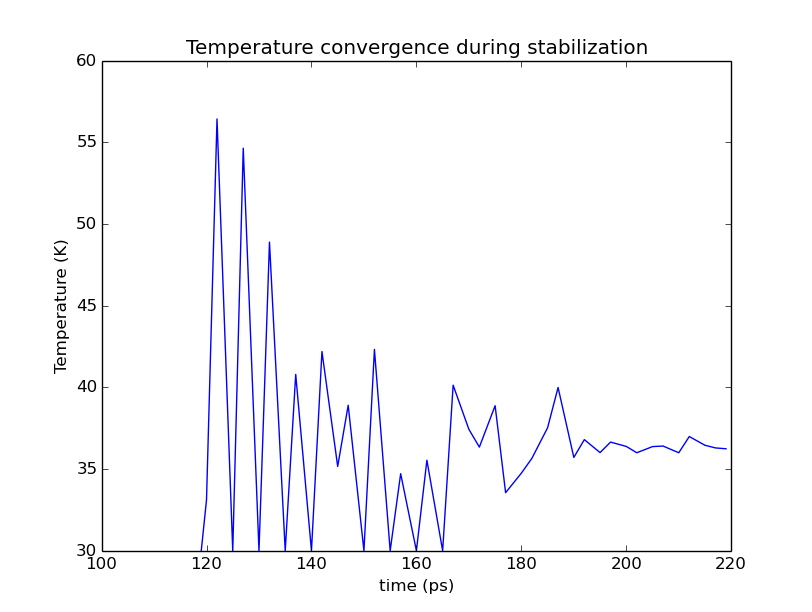}
\end{center}
\subsection{Conclusion}
Therefore, several MD simulations have been done to simulate the effect of precipitate on matrix mechanical properties. The generation of stacking faults and misfits have been investigated and the subsequent strengthening effect has been explained reasonably well, within the confines of the simulation box, by plotting stress-strain response. The strength has been shown proportional to particle size and inter-particle distance. This, to an extent, explains precipitation hardening. 

\subsection{Outlook}
There exists a specific relationship between SFW (Stacking fault width) and inter-particle distance \cite{zhu1995macroscopic}. It remains to be seen if the same can be verified by the MD simulation. Plus the critical particle distance, below which the dislocation just cuts through the particle hasn't been investigated very well. It is thought to cause a softening in the material. A relationship can be established between particle size, volume fraction, shape and critical dislocation cut-through distance. 
\section{Acknowledgments}
We would first like to thank Prof. Anand Kanjarla for his guidance and motivation. We would also like to thank A.R.G.Sreekar and Abhishek Shandilya for sharing their work with us. We would like to extend our warm regards to Sri Hari (T.A for the course), for helping us through odd hours. 
The availability of high-performance computing facility GNR has also been greatly helpful.
\bibliographystyle{plain}
\bibliography{GAUtham.bib}

\end{document}